\documentclass[10pt,aps,prd,preprintnumbers,showpacs,nofootinbib,superscriptaddress,notitlepage]{revtex4-1}

\usepackage{graphicx}
\usepackage{color}
\usepackage{rotating}
\usepackage{amsmath}
\usepackage{amssymb}
\usepackage{enumerate}
\usepackage{multirow}
\usepackage[colorlinks=true,citecolor=blue,urlcolor=blue]{hyperref}
\usepackage{subcaption}

\def\be{\begin{equation}}
\def\ee{\end{equation}}
\def\bea{\begin{eqnarray}}
\def\eea{\end{eqnarray}}

\def\gev{\, {\rm GeV}}
\def\mev{\, {\rm MeV}}

\newcommand{\gsim}{\lower.7ex\hbox{$\;\stackrel{\textstyle>}{\sim}\;$}}
\newcommand{\lsim}{\lower.7ex\hbox{$\;\stackrel{\textstyle<}{\sim}\;$}}

\newcommand{\cm}{\rm cm}
\newcommand{\sr}{\rm sr}
\newcommand{\s}{\rm s}
\newcommand{\yr}{\rm yr}

\begin{document}

\setlength{\baselineskip}{0.25in}

\setcounter{footnote}{0}
\setcounter{page}{2}
\setcounter{figure}{0}
\setcounter{table}{0}


\title{Indirect detection of low-mass dark matter 
through the $\pi^0$ and $\eta$ windows}

\author{J.~G.~Christy}
\affiliation{\mbox{Department of Physics \& Astronomy,
University of Hawai'i, Honolulu, HI 96822, USA}}

\author{Jason Kumar}
\affiliation{\mbox{Department of Physics \& Astronomy,
University of Hawai'i, Honolulu, HI 96822, USA}}

\author{Arvind Rajaraman
}
\affiliation{\mbox{Department of Physics and Astronomy, University of California, Irvine, CA 92697, USA}}

\preprint{ UCI-TR-2022-08, UH511-1328-2022}

\begin{abstract}

We consider the search for gamma rays produced by the annihilation or decay of low-mass dark matter 
which couples to quarks.  In this scenario, most of the photons are produced from the decays of 
$\pi^0$ or $\eta$ mesons.  These decays produce distinctly different photon signatures due to the 
difference in meson mass.  We assess the ability of the future MeV-range observatories to 
constrain the hadronic final states produced by dark matter annihilation or decay 
from the shape of the resulting photon spectrum.  We then comment on how this information can 
be used to determine properties of the dark matter coupling to the quark current, based on 
the approximate symmetries of low-energy QCD.

\end{abstract}

\maketitle


\section{Introduction}

The annihilation or decay of dark matter to Standard 
Model particles can have interesting features if the 
dark matter is relatively light 
($m_{\chi} \lesssim {\cal O}(\gev)$), particularly if 
dark matter couples to quarks (see, for example,~\cite{Boddy:2015efa,Boddy:2015fsa,Boddy:2016fds,Bartels:2017dpb,Kumar:2018heq,Berger:2019aox,Coogan:2019qpu,Coogan:2021rez,Coogan:2021sjs,Reimitz:2021nsi,Plehn:2021sam}).  Because the dark matter 
mass is not very much heavier than that of the 
lightest hadrons, one must consider these processes 
as interactions between dark matter and mesons.   
The dominant gamma-ray signature of these processes 
then arises from the decay of these light mesons.

The main mechanisms by which the lightest pseudoscalar 
and vector mesons produce photons are the decays 
$\pi^0, \eta \rightarrow \gamma \gamma$, where the 
$\pi^0$ and  $\eta$ are either produced directly from dark 
matter annihilation/decay, or as the subsequent decay 
products of other heavier mesons.
Photons are also produced by the decays of the $\eta'$ and $\omega$, but they  
are considerably heavier, and not always kinematically accessible for 
low-energy processes.
Because of the 
low center-of-mass energy of the process, the 
$\pi^0$ and $\eta$ are typically produced with only 
moderate boost.  As a result, the photon signals from 
the $\pi^0$ and $\eta$ are relatively easy to distinguish 
from each other.  Our goal in this paper is to study 
the ability of future MeV-range gamma-ray experiments, 
such as e-ASTROGAM~\cite{e-ASTROGAM:2017pxr} or AMEGO~\cite{Kierans:2020otl}, 
to distinguish between 
the possible final states produced by dark matter 
annihilation/decay, based on the differences in the 
photon spectral shape arising from the expected numbers 
of $\pi^0$ and $\eta$ produced per interaction.

These results are of relevance because the relative 
number of $\pi^0$ and $\eta$ produced in each 
interaction is controlled by the Lorentz and isospin 
structure of the interaction.  For example, if dark 
matter couples to light quarks through a 
vector interaction, then one expects a small  
amount of $\eta$ production~\cite{Berger:2019aox}.  On the other hand, 
if the coupling structure is scalar or 
pseudoscalar, then a significant fraction of $\eta$s 
can be produced, depending on the isospin structure of 
the interaction~\cite{Kumar:2018heq}.  Thus, determining the meson content of 
the final states produced by dark matter decay or annihilation 
can reveal information about the symmetry structure of 
microscopic dark matter interactions.

It is important to emphasize that these results do 
not depend on the details or validity of 
chiral perturbation 
theory, but rather on the approximate symmetries of 
quark interactions at energies well below the electroweak 
scale.  If we assume that dark matter interactions with 
quarks respect $C$ and $P$, then the final state quantum 
numbers under $J$, $C$, $P$ and isospin will be the same 
as those of the quark current to which the dark matter 
couples.  At the energies which we consider, there 
may be large corrections to amplitudes 
computed using the 
chiral Lagrangian, but these will not affect our 
results, which depend only on the final states 
which are allowed by the symmetries of the low-energy 
theory.

We will see that, given the energy resolution, angular 
resolution, and exposure expected from the next generation 
of MeV-range gamma-ray experiments, one would expect to 
be able to distinguish final states in which an $\eta$ 
is produced.  It is much more difficult to distinguish final 
states in which only pions are produced.  Even an increase 
in exposure by a factor of 10 is not sufficient to clearly 
distinguish final states which only involve pions.  But improvement in 
the expected energy resolution would allow one to distinguish between 
final states which only produce pions.

The plan of this paper is as follows.  We will discuss the 
photon spectrum in Sec.~\ref{sec:Spectrum}.  We will 
describe our analysis and results in Sec.~\ref{sec:AnalysisResults}. 
We discuss the implications of these results in Sec.~\ref{sec:Discussion}.  Finally, 
we conclude in Sec.~\ref{sec:Conclusion}.

\section{Gamma-ray spectrum}
\label{sec:Spectrum}

We will focus on neutral mesonic final states which contain at most two or three mesons.
We expect that, provided at least one such state is kinematically accessible and 
allowed by the approximate symmetries of the theory, it should dominate over phase 
space suppressed final states with larger numbers of mesons.  Since we are interested in 
final states which respect the approximate symmetries of QCD, we are only interested 
in states with vanishing net strangeness.

In Table 1, we list all of the light pseudoscalar and vector mesons, as well 
as the branching fractions to decay channels which produce non-negligible numbers of 
photons~\cite{Zyla:2020zbs}.\footnote{Note, we include the $\pi^\pm$ and $\rho^0$ for completeness, 
despite the fact that all of their decay channels with significant branching fraction 
produce only a small number of photons.}  It is readily seen that none of the mesons which are lighter than the 
$\eta'$ has a significant branching fraction to states which contain an $\eta$.  
As a result, the only primary final states 
for which photons are produced by $\eta$ decay are $\pi^0 \eta$ and $\pi \pi \eta$.  For 
all other states, photons are produced almost entirely from $\pi^0$, where the 
$\pi^0$ is produced either in the primary process or in the cascade decays of heavier 
mesons.

The photon spectrum produced by isotropic diphoton decay has been discussed 
in detail in Refs.~\cite{Boddy:2016hbp,Berger:2019aox}.  
Of particular relevance for this work is that, if the 
parent particle, with mass $m$, is not heavily boosted, then the photon spectrum is 
relatively tightly peaked at $E_* = m/2$.  Indeed, if plotted against $\log E_\gamma$, 
it can be shown that the photon spectrum has a global maximum at $E_*$.
Since the $\eta$ is $\sim 4$ times heavier 
than the $\pi^0$ ($m_{\pi^0} = 135~\mev$, 
$m_\eta = 548~\mev$), this implies that final states containing an $\eta$ will yield  
photon spectra with support at higher energies than those of final states 
involving only $\pi^0$.  This 
feature will be useful in allowing us to distinguish the photon spectra arising from 
different final states, based on their differing $\eta$ content.

\begin{table}[h!]
  \begin{center}
    \begin{tabular}{|c|c|c|c|c|}
      \hline
      Meson & Mass (\mev)& $J^{P(C)}$ & Decay mode & BF\\
      \hline\hline
      \multirow{1}{*}{$\pi^0$}
      & 135 & $0^{-(+)}$ & $2\gamma$ & 98.8\%\\
      \hline
      \multirow{1}{*}{$\pi^\pm$}
	  & 140 & $0^{-}$ & $\mu^\pm \nu$& 100\%\\
      \hline
      \multirow{4}{*}{$K^\pm$}&\multirow{4}{*}{494}&\multirow{4}{*}{$0^-$}
      & $\pi^\pm\pi^0$& 20.7\%\\
        &&& $\pi^0e^\pm\nu$ & 5\%\\
	    &&& $\pi^0\mu^\pm\nu$ & 3.4\%\\
		&&& $\pi^\pm\pi^0\pi^0$& 1.8\%\\
	    \hline
      \multirow{1}{*}{$K^0_S$}
		& 498 & $0^-$ & $\pi^0\pi^0$& 30.7\%\\
      \hline
		\multirow{2}{*}{$K^0_L$}&\multirow{2}{*}{498}&\multirow{2}{*}{$0^-$}
		& $\pi^0\pi^0\pi^0$& 19.5\%\\
		&&& $\pi^+\pi^-\pi^0$&  12.5\%\\
      \hline
      \multirow{3}{*}{$\eta$}
	  &&& $2\gamma$& 39.4\%\\
	  & 548 & $0^{-(+)}$ & $3\pi^0$& 32.7\%\\
	  &&& $\pi^+\pi^-\pi^0$& 22.9\%\\
	  \hline
    \end{tabular}
    \hspace{4pt}
    \begin{tabular}{|c|c|c|c|c|}
      \hline
      Meson & Mass (\mev)& $J^{P(C)}$ & Decay mode & BF\\
      \hline\hline
      \multirow{1}{*}{$\rho^\pm$}
	    & 775 & $1^{-(-)}$ & $\pi^\pm \pi^0$& 100\%\\
      \hline
      \multirow{1}{*}{$\rho^0$}
		& 775 & $1^{-(-)}$ & $\pi^+\pi^-$ & 100\%\\
      \hline
      \multirow{2}{*}{$\omega$}&\multirow{2}{*}{783}&\multirow{2}{*}{$1^{-(-)}$}
		& $\pi^+\pi^-\pi^0$& 89\%\\
		&&& $\pi^{0}\gamma$& 8\%\\
      \hline
		\multirow{1}{*}{$K^{*0}$}
		& 892 & $1^-$ & $K\pi$& 100\%\\
      \hline
		\multirow{1}{*}{$K^{*\pm}$}
		& 892 & $1^-$ & $K\pi$& 100\%\\
      \hline
      \multirow{3}{*}{$\eta'$}
	  &&& $\pi^+\pi^-\eta$& 42.5\%\\
	  & 958 & $0^{-(+)}$ & $\rho^0\gamma$& 29.5\%\\
	  &&& $\pi^0\pi^0\eta$& 22.4\%\\
      \hline
      \multirow{3}{*}{$\phi$}
		&&& $K^+K^-$& 49.2\%\\
		& 1019 & $1^{-(-)}$ & $K^0_LK^0_S$& 34\%\\
		&&& $\rho\pi+\pi^+\pi^-\pi^0$& 15.2\%\\
      \hline
    \end{tabular}
		    \caption{The relevant masses, 
		    $J^{P(C)}$ quantum numbers, decay modes and branching fractions (BF) for the light pseudoscalar and vector mesons.
		    }
\label{Tab:BranchingFraction}
  \end{center}
\end{table}

\begin{figure}[h]
\begin{subfigure}[b]{0.49\textwidth}
\centering
\includegraphics[width=\textwidth]{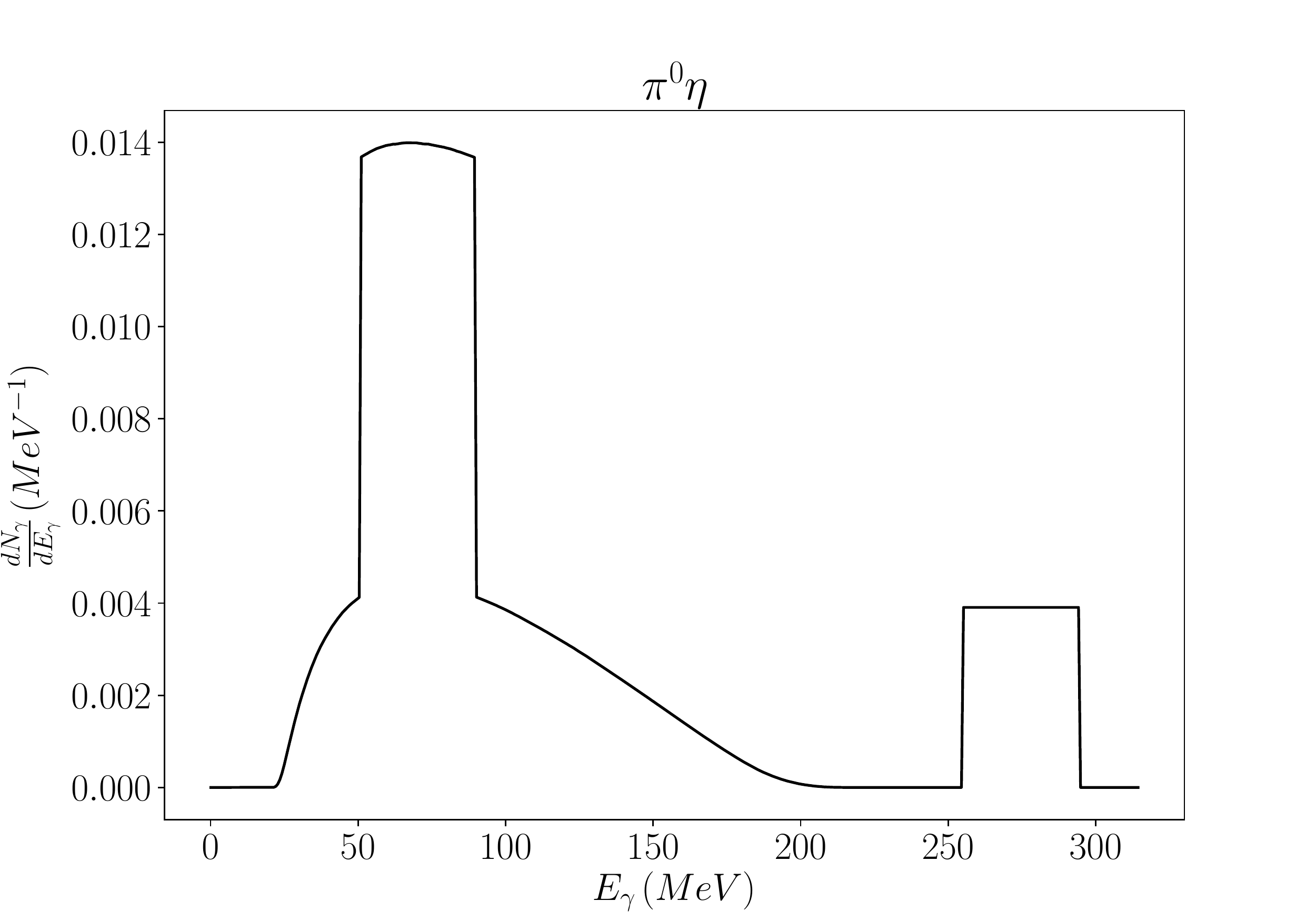}
\end{subfigure}
\begin{subfigure}[b]{0.49\textwidth}
\centering
\includegraphics[width=\textwidth]{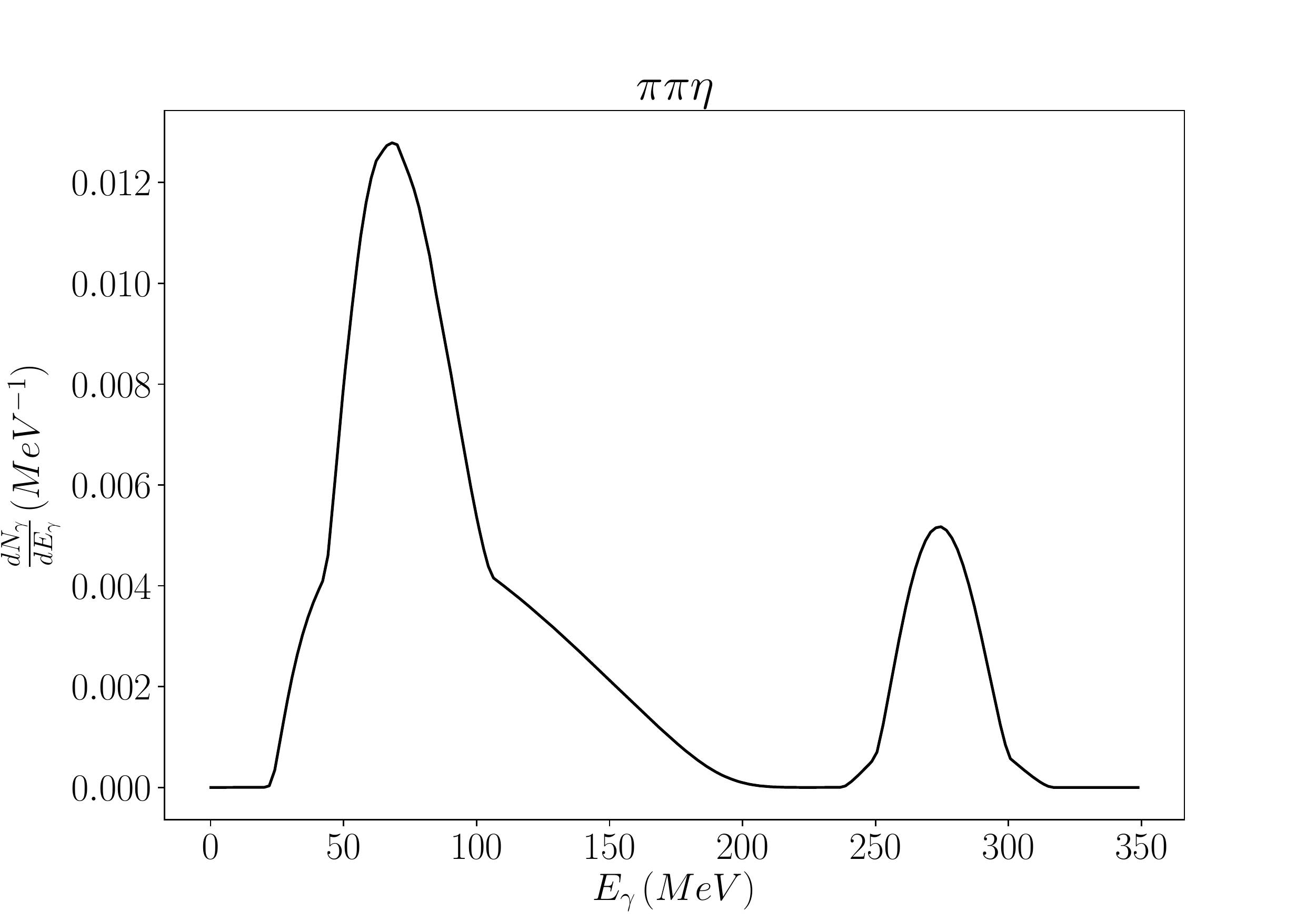}
\end{subfigure}
\begin{subfigure}[b]{0.49\textwidth}
\centering
\includegraphics[width=\textwidth]{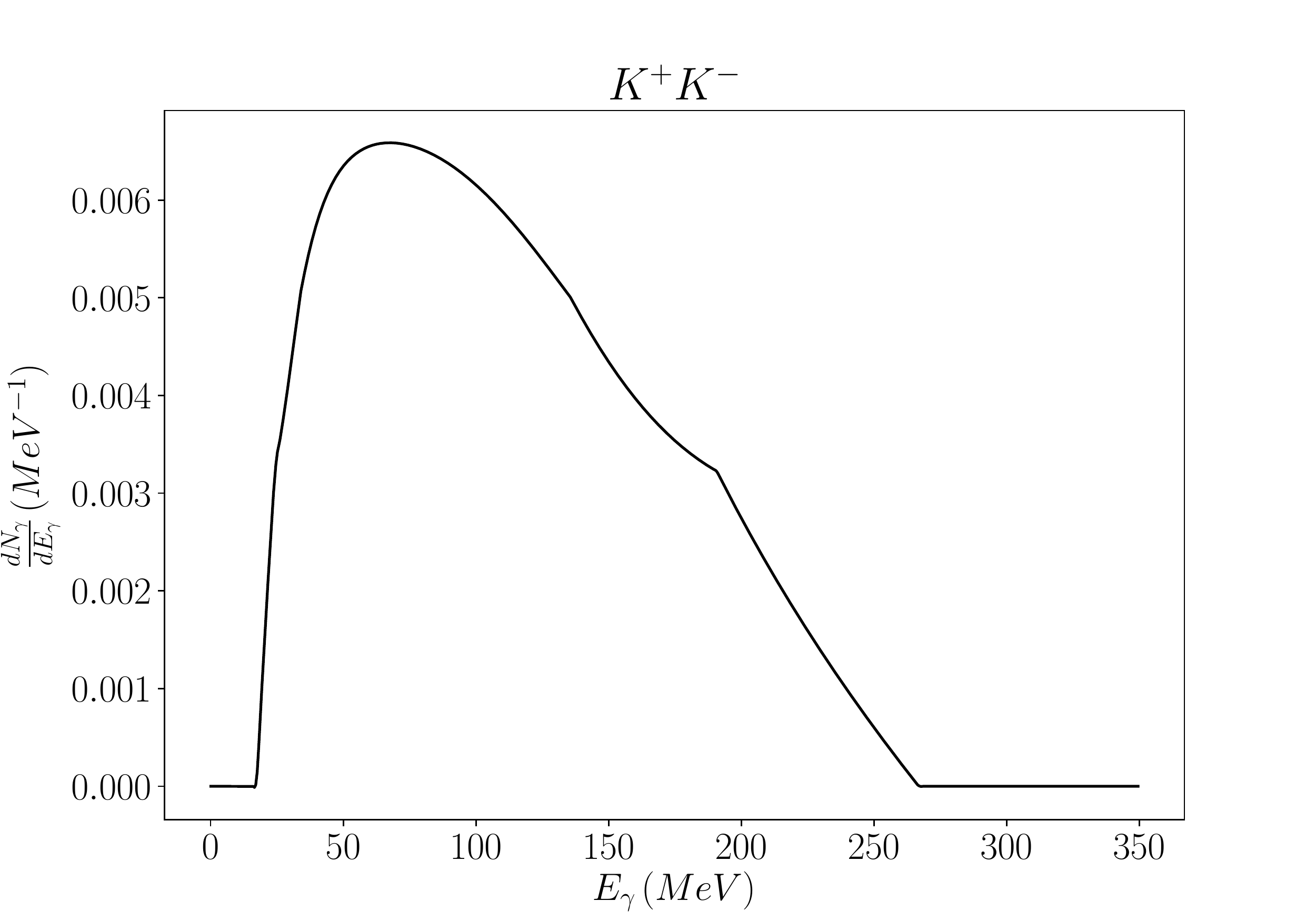}
\end{subfigure}
\begin{subfigure}[b]{0.49\textwidth}
\centering
\includegraphics[width=\textwidth]{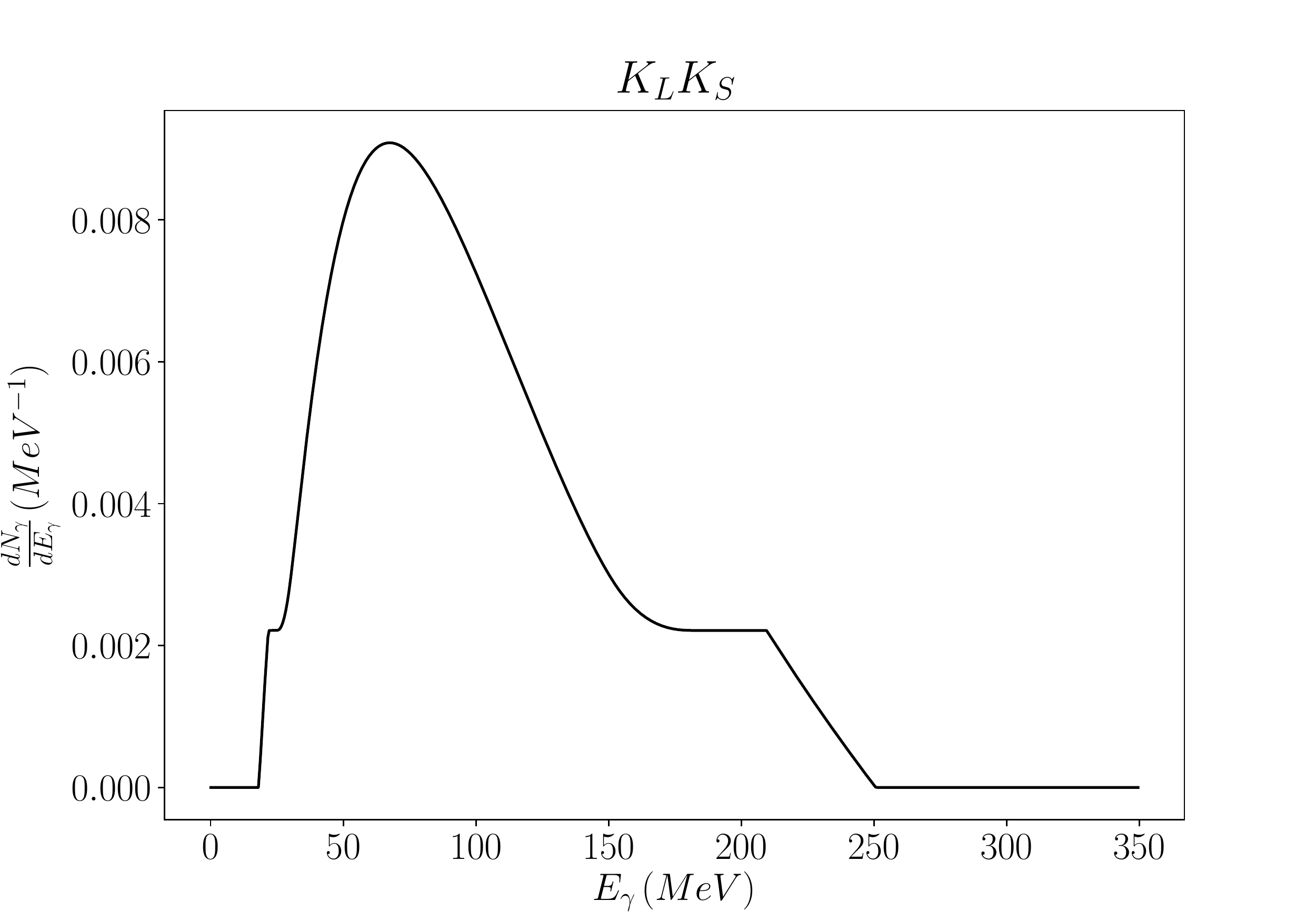}
\end{subfigure}
\caption{The gamma ray spectra for considered decay channels near their respective thresholds. The top left panel shows the spectrum for $\pi^0 \eta$ with $\sqrt{s} = 690$ \mev. The top right panel shows the spectrum for $\pi \pi \eta$ with $\sqrt{s} = 835$ MeV. The bottom left panel shows the spectrum for $K^+ K^-$ with $\sqrt{s} = 1000$ \mev. The bottom right panel shows the spectrum for $K_L K_S$ with $\sqrt{s} = 1000$ \mev. All four spectra have peaks around $m_\pi /2$. This peak is narrower when the decay products include $\eta$s; such states also feature a secondary peak near $m_\eta /2$.}
\label{fig:spectra}
\end{figure}

To illustrate, we consider four particular final states: 
$\pi^0 \eta$, $\pi \pi \eta $ (with the $\pi \pi$ state having isospin 0), 
$K^+ K^-$ and $K_L K_S$.  
We plot the photon energy 
spectrum (normalized to unity) for the $\pi^0 \eta$, 
$\pi \pi \eta$, $K^+ K^-$ and $K_L K_S$ states (with 
$\sqrt{s}$ (MeV) $= 690, 835, 1000,$ and $1000$ respectively), in Fig.~\ref{fig:spectra}.  As can be seen 
by eye, the most marked similarities and differences in the spectra 
are related to the $\eta$ content of the final state.  States with 
an $\eta$ produce a narrow peak near $m_\eta /2$ and another narrow 
peak near $m_\pi /2$, while states with no $\eta$s produce a single 
broader peak near $m_\pi /2$.  The width of these features is determined by 
how boosted the $\eta$ and $\pi^0$ are, in the center-of-mass frame.  As a 
result, these differences will become 
less significant at larger center-of-mass energies.

\begin{figure}[h]
\centering
\begin{subfigure}[b]{0.32\textwidth}
\centering
\includegraphics[width=\textwidth]{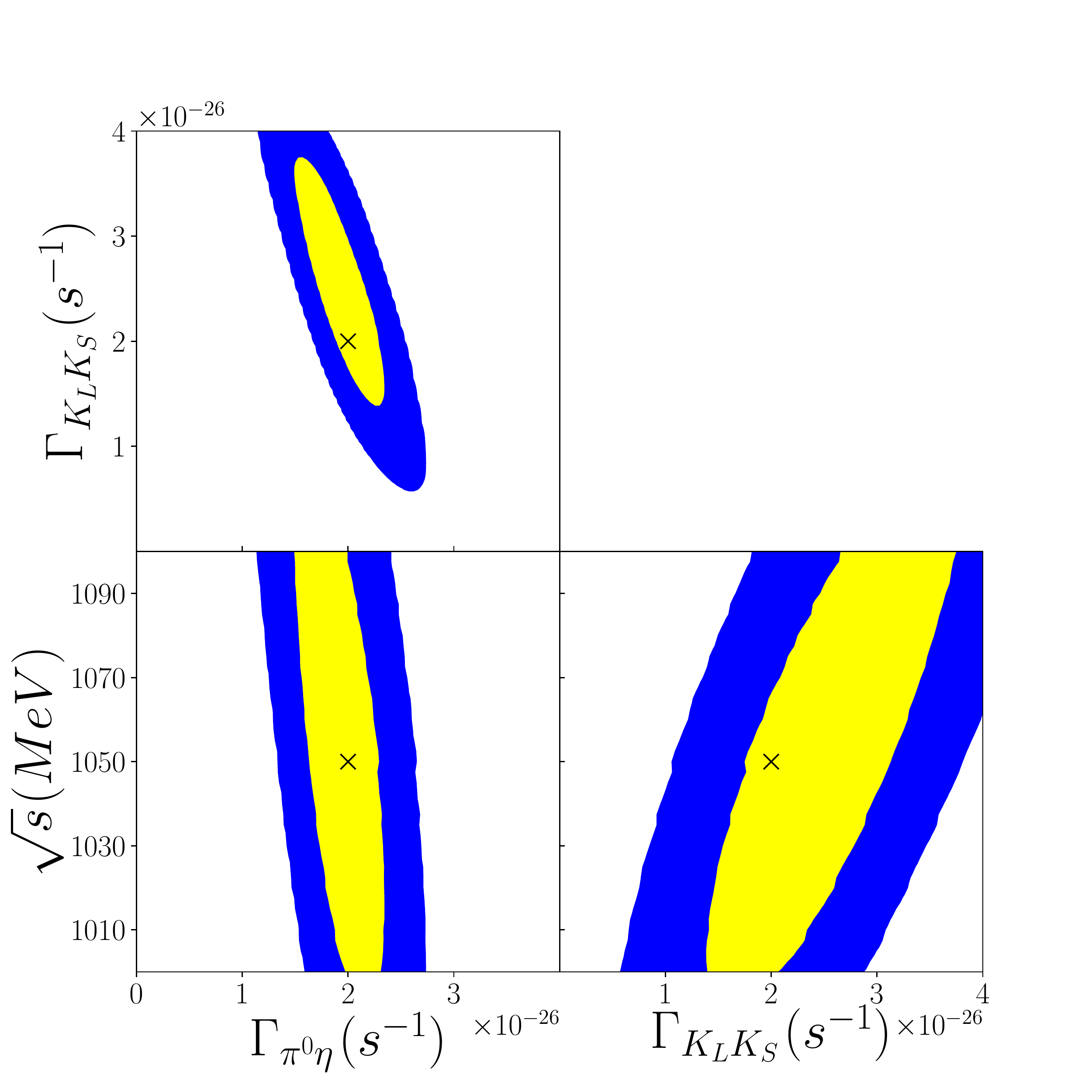}
\end{subfigure}
\begin{subfigure}[b]{0.32\textwidth}
\centering
\includegraphics[width=\textwidth]{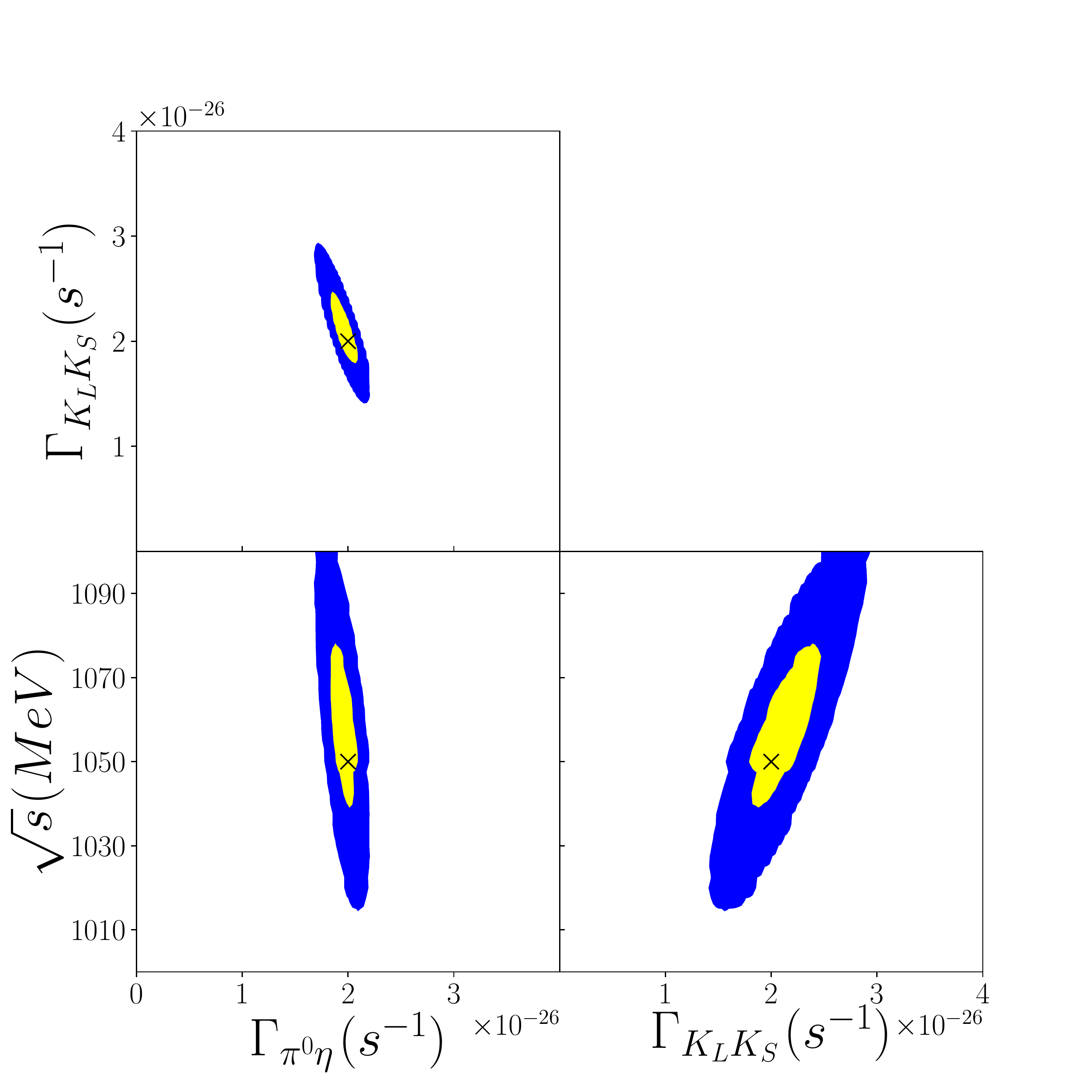}
\end{subfigure}
\begin{subfigure}[b]{0.32\textwidth}
\centering
\includegraphics[width=\textwidth]{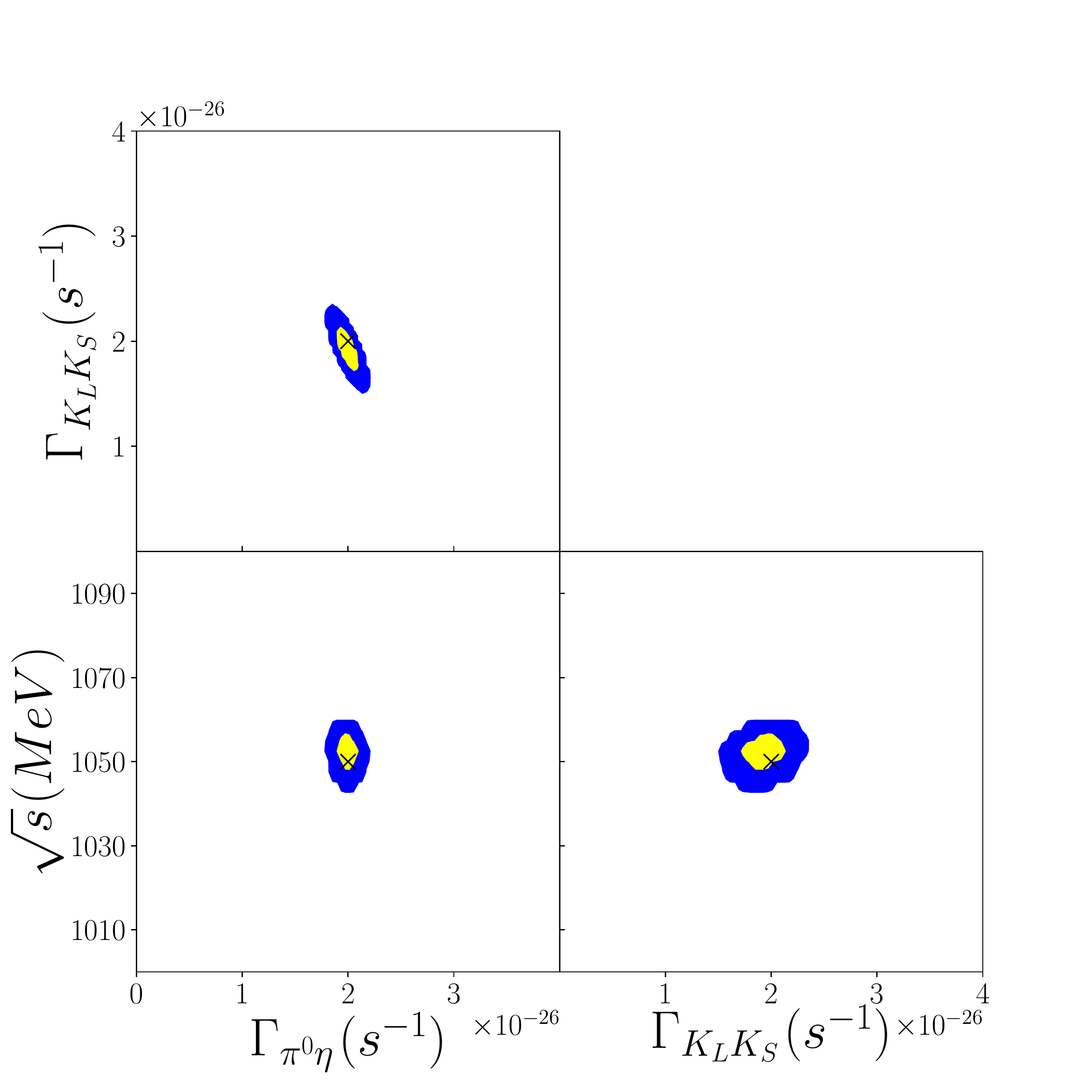}
\end{subfigure}
\caption{Projections of three-dimensional 2$\sigma$ (yellow) and $5\sigma$ (blue) confidence level contours onto 
two-dimensional subspaces, as labeled by the axes, for the $\pi\eta$ and $K_LK_S$ decay channels. The black X represents the model with which the mock data were generated. The left panel shows 3000 $\cm^2~\yr$ exposure and 30\% energy resolution. The middle panel shows 30000 $\cm^2~\yr$ exposure and 30\% energy resolution. The right panel shows 30000 $\cm^2~\yr$ exposure and 3\% energy resolution.
\label{fig:PiEtaKLKS}
}
\end{figure}

\begin{figure}[h]
\centering
\begin{subfigure}[b]{0.32\textwidth}
\centering
\includegraphics[width=\textwidth]{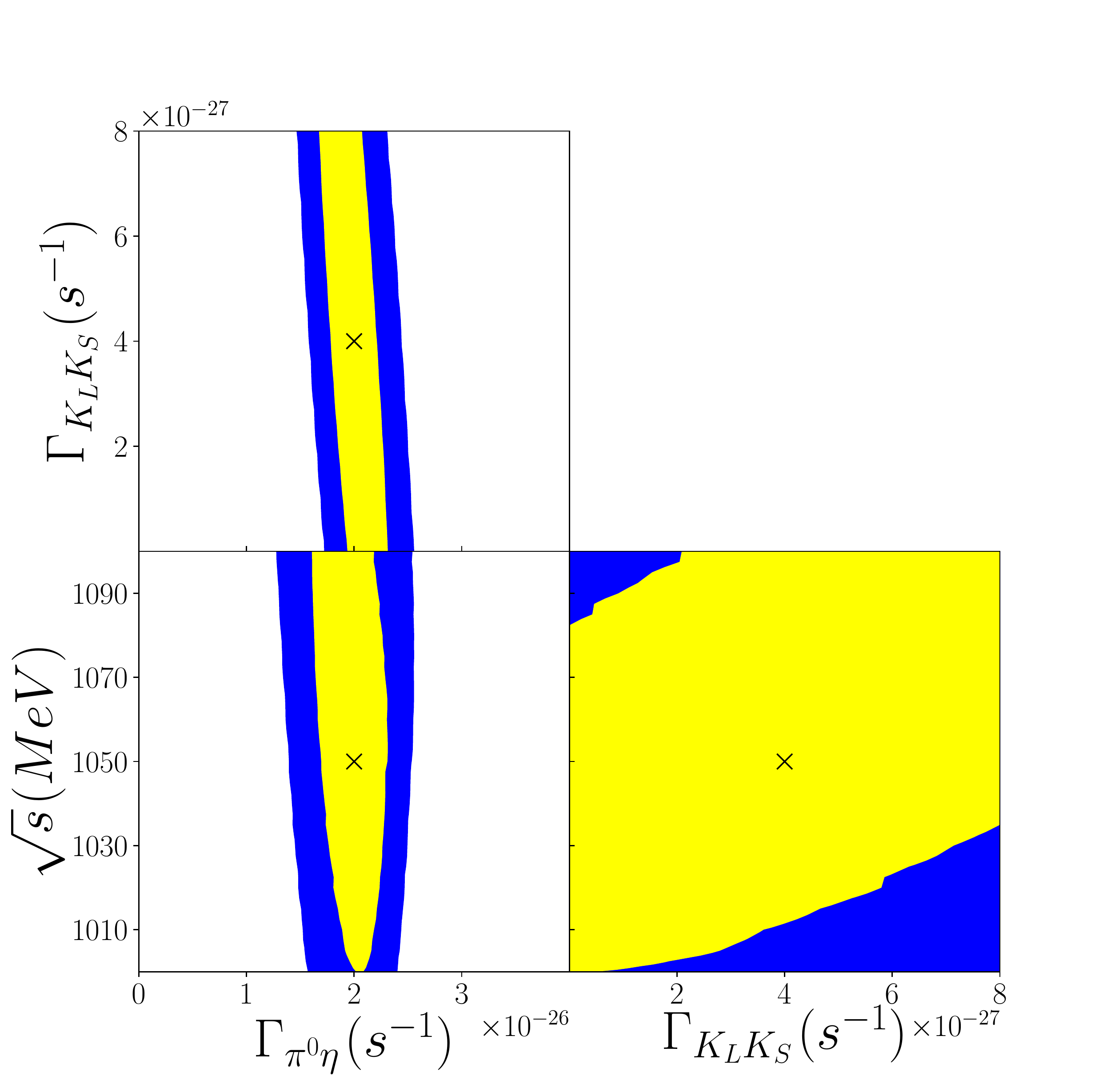}
\end{subfigure}
\begin{subfigure}[b]{0.32\textwidth}
\centering
\includegraphics[width=\textwidth]{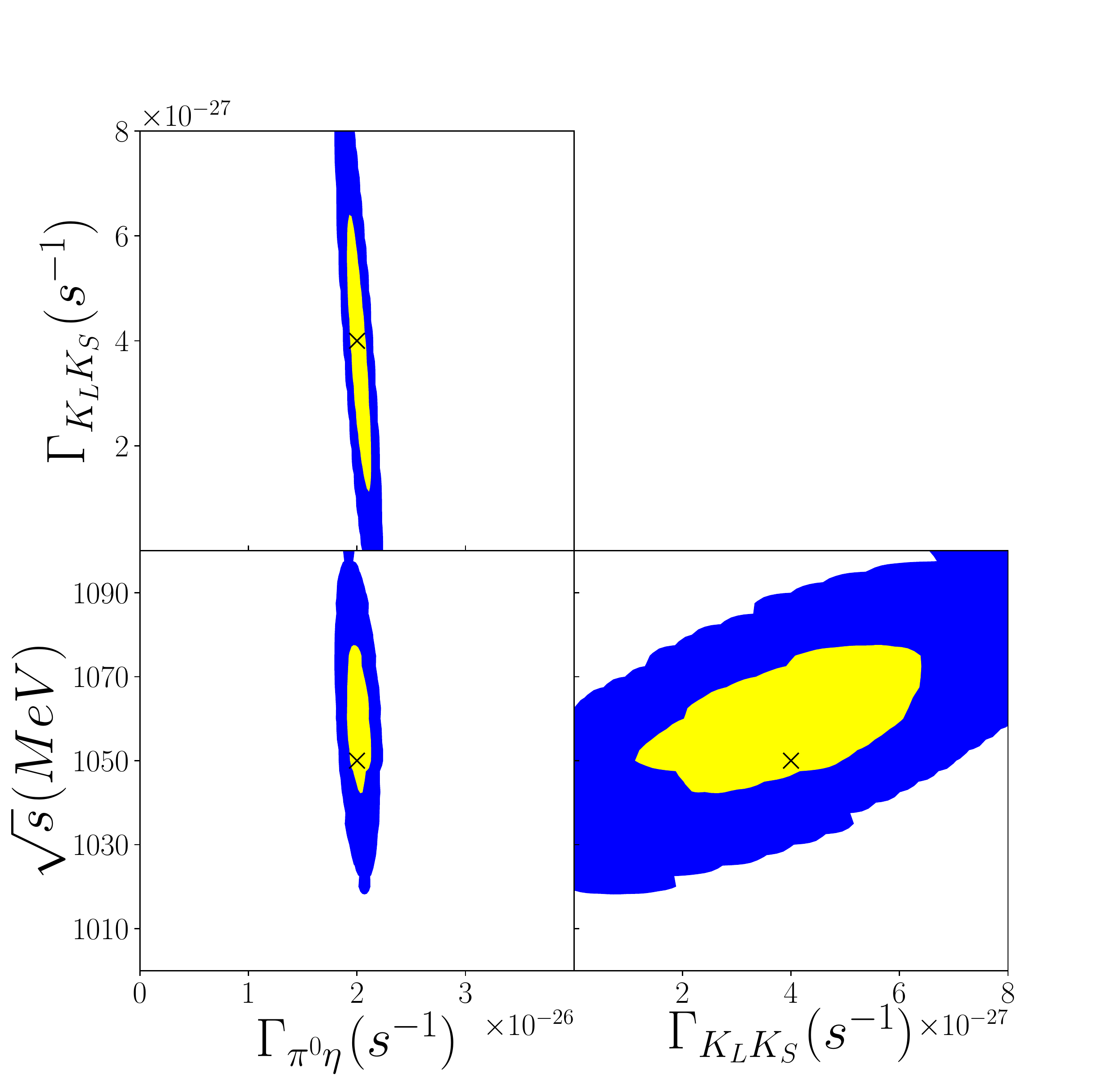}
\end{subfigure}
\begin{subfigure}[b]{0.32\textwidth}
\centering
\includegraphics[width=\textwidth]{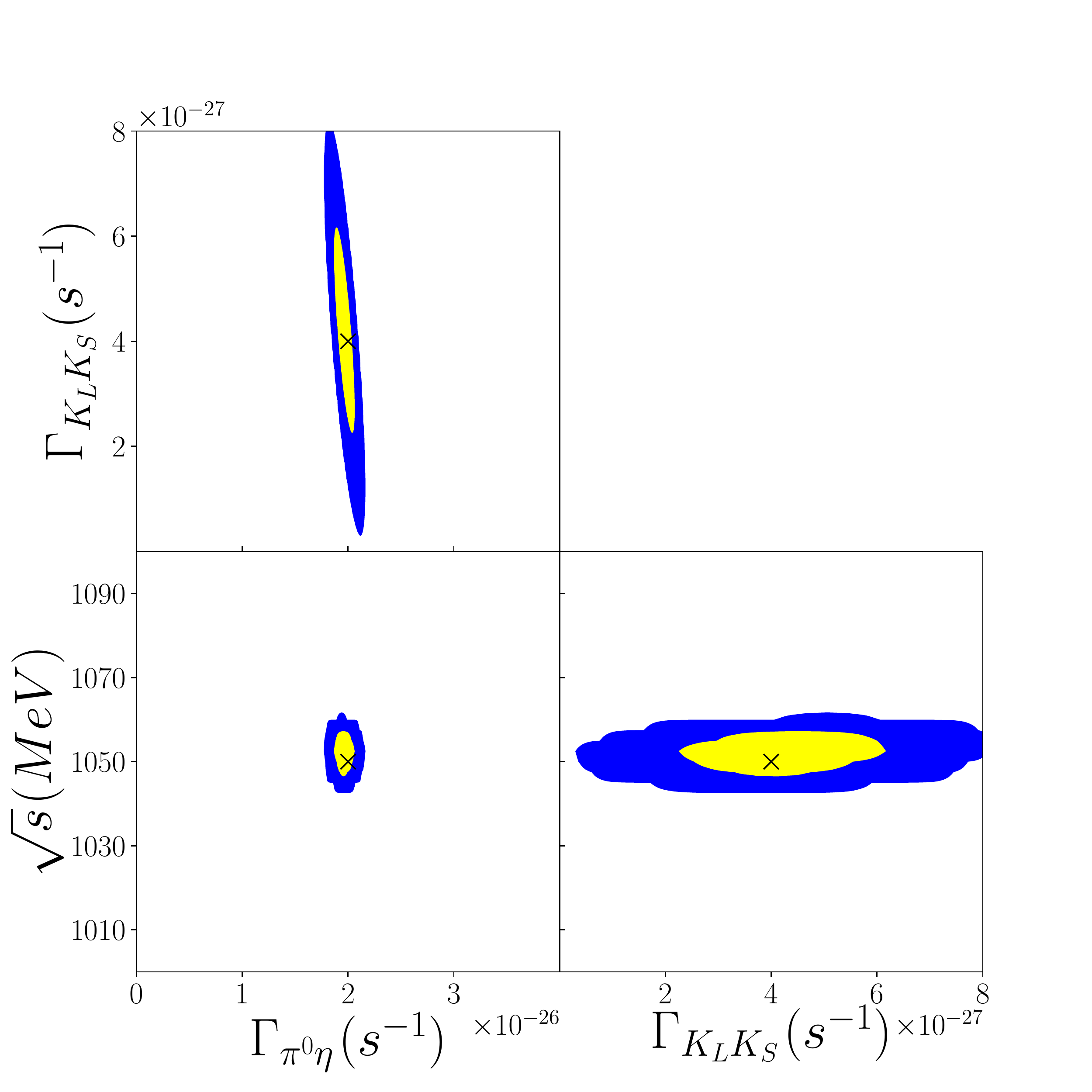}
\end{subfigure}
\caption{Similar to Fig.~\ref{fig:PiEtaKLKS}, except the true model has 
$\Gamma_{\pi^0 \eta} = 2 \times 10^{-26}~\s^{-1}$, $\Gamma_{K_L K_S} = 4 \times 10^{-27}~\s^{-1}$.
\label{fig:PiEtaKLKS_phase_space_corrected}
}
\end{figure}

\begin{figure}[h]
\centering
\begin{subfigure}[b]{0.32\textwidth}
\centering
\includegraphics[width=\textwidth]{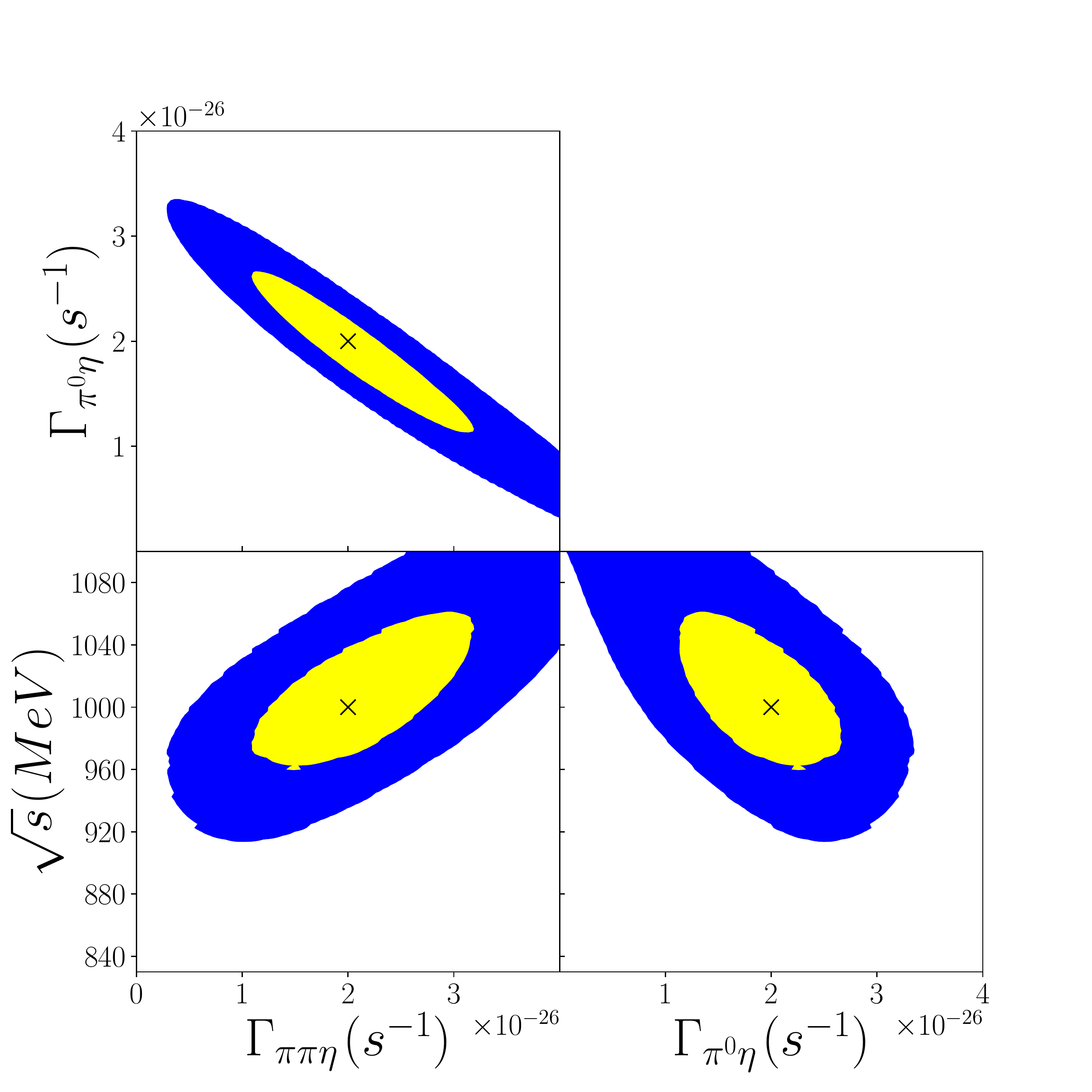}
\end{subfigure}
\begin{subfigure}[b]{0.32\textwidth}
\centering
\includegraphics[width=\textwidth]{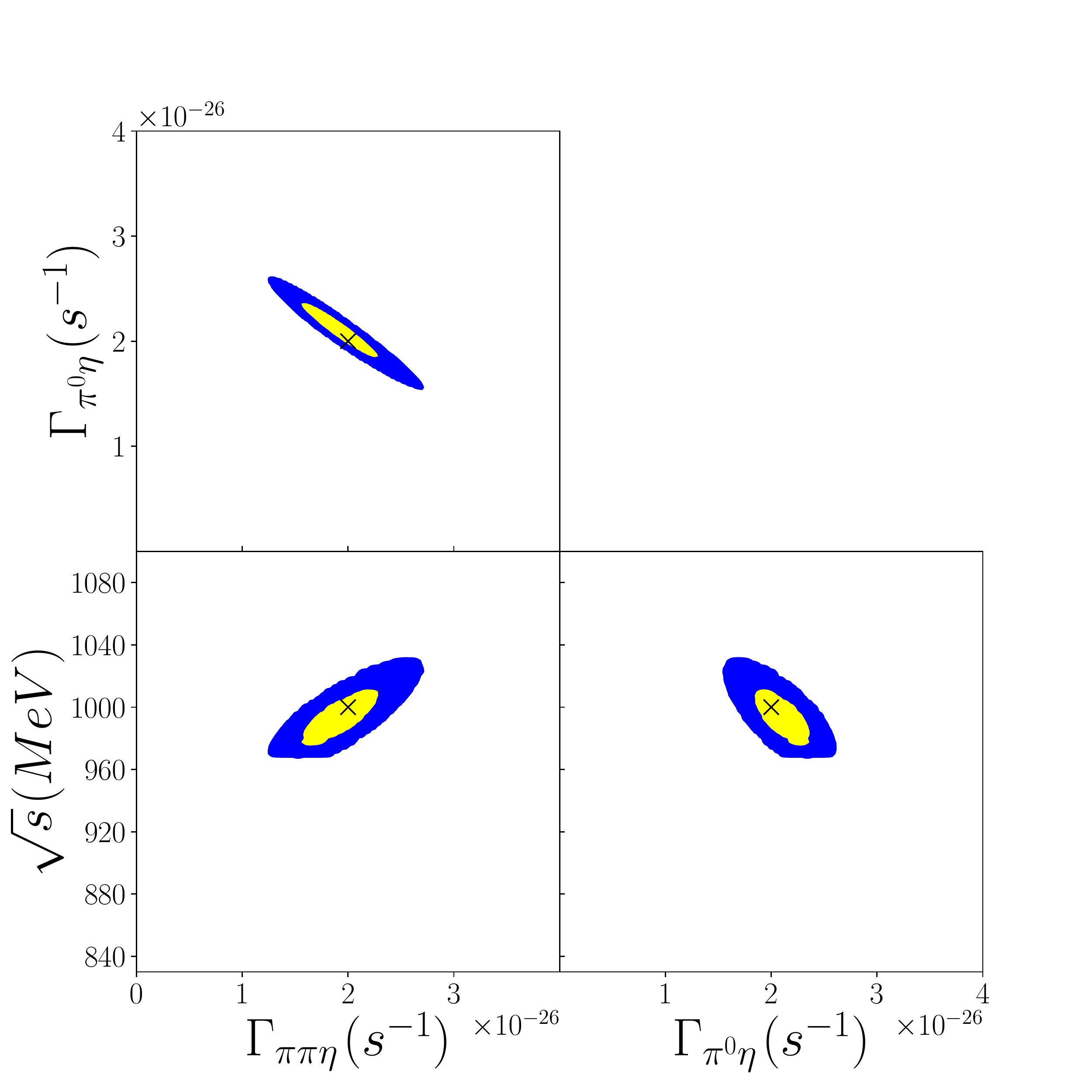}
\end{subfigure}
\begin{subfigure}[b]{0.32\textwidth}
\centering
\includegraphics[width=\textwidth]{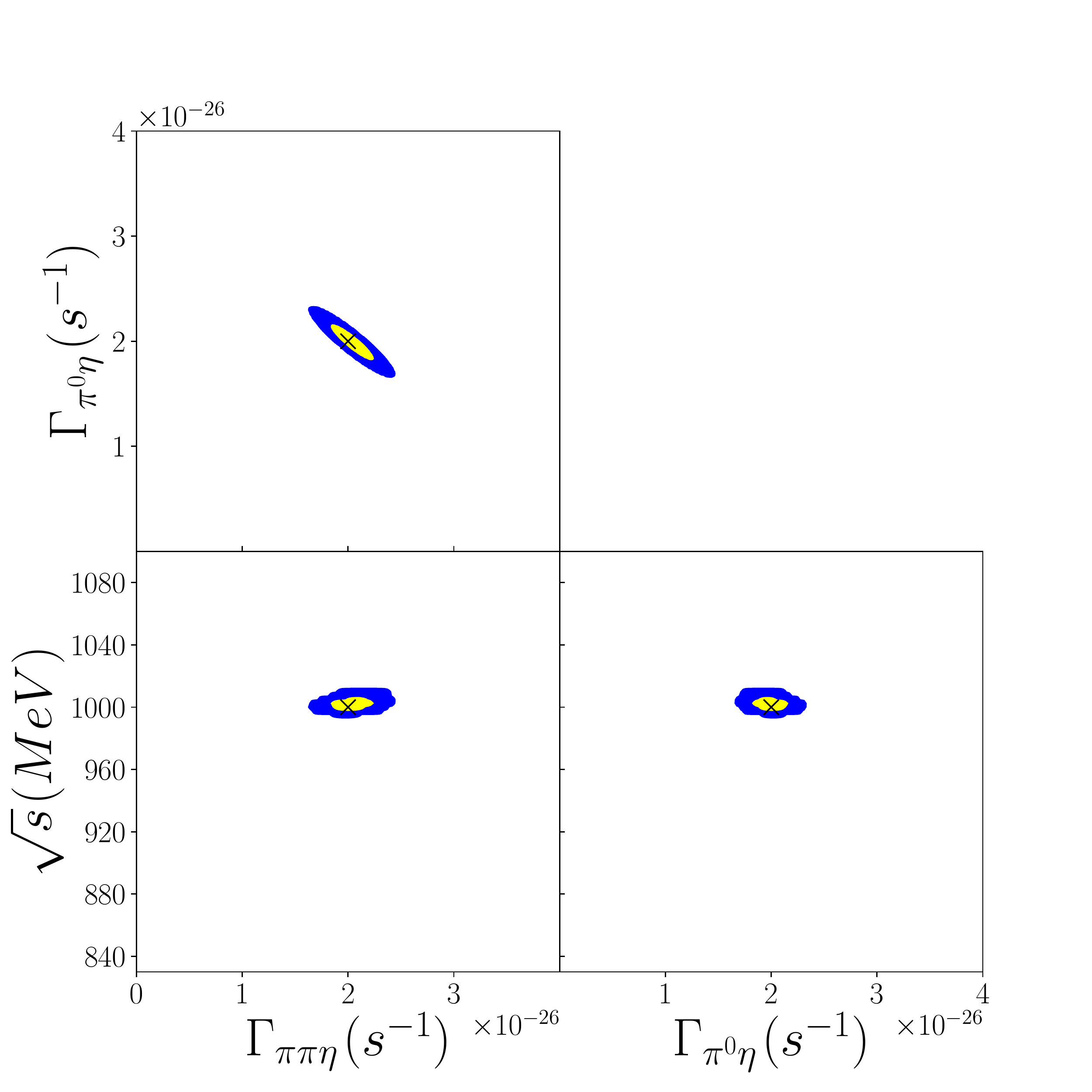}
\end{subfigure}
\caption{Projections of three-dimensional 2$\sigma$ (yellow) and $5\sigma$ (blue) confidence level contours onto 
two-dimensional subspaces, as labeled by the axes, for the $\pi\pi\eta$ and $\pi\eta$ decay channels. The black X represents the model with which the mock data was generated. The left panel shows 3000 $\cm^2~\yr$ exposure and 30\% energy resolution. The middle panel shows 30000 $\cm^2~\yr$ exposure and 30\% energy resolution. The right panel shows 30000 $\cm^2~\yr$ exposure and 3\% energy resolution.
\label{fig:PiEtaPiPiEta_30}
}
\end{figure}

\begin{figure}[h]
\centering
\begin{subfigure}[b]{0.32\textwidth}
\centering
\includegraphics[width=\textwidth]{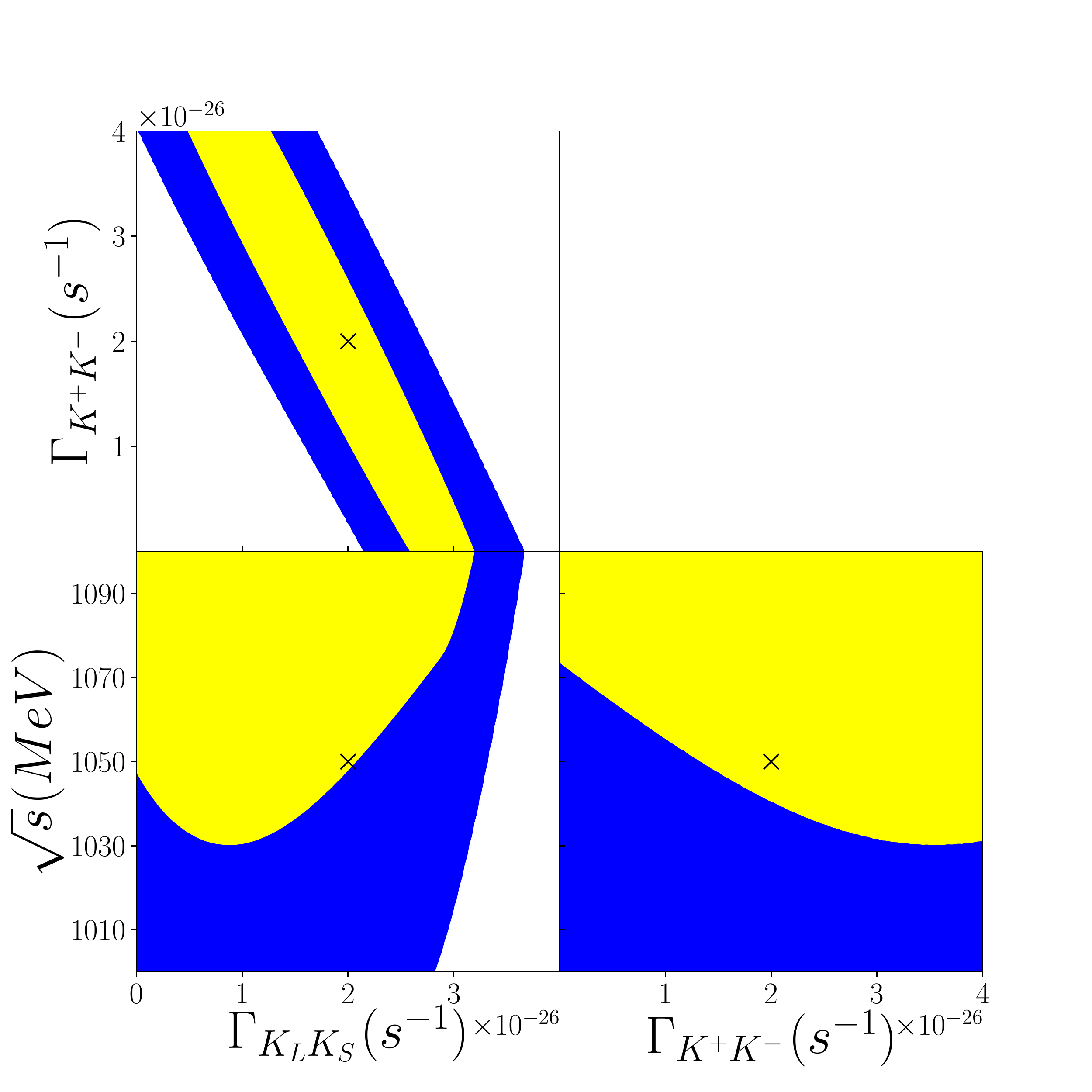}
\end{subfigure}
\begin{subfigure}[b]{0.32\textwidth}
\centering
\includegraphics[width=\textwidth]{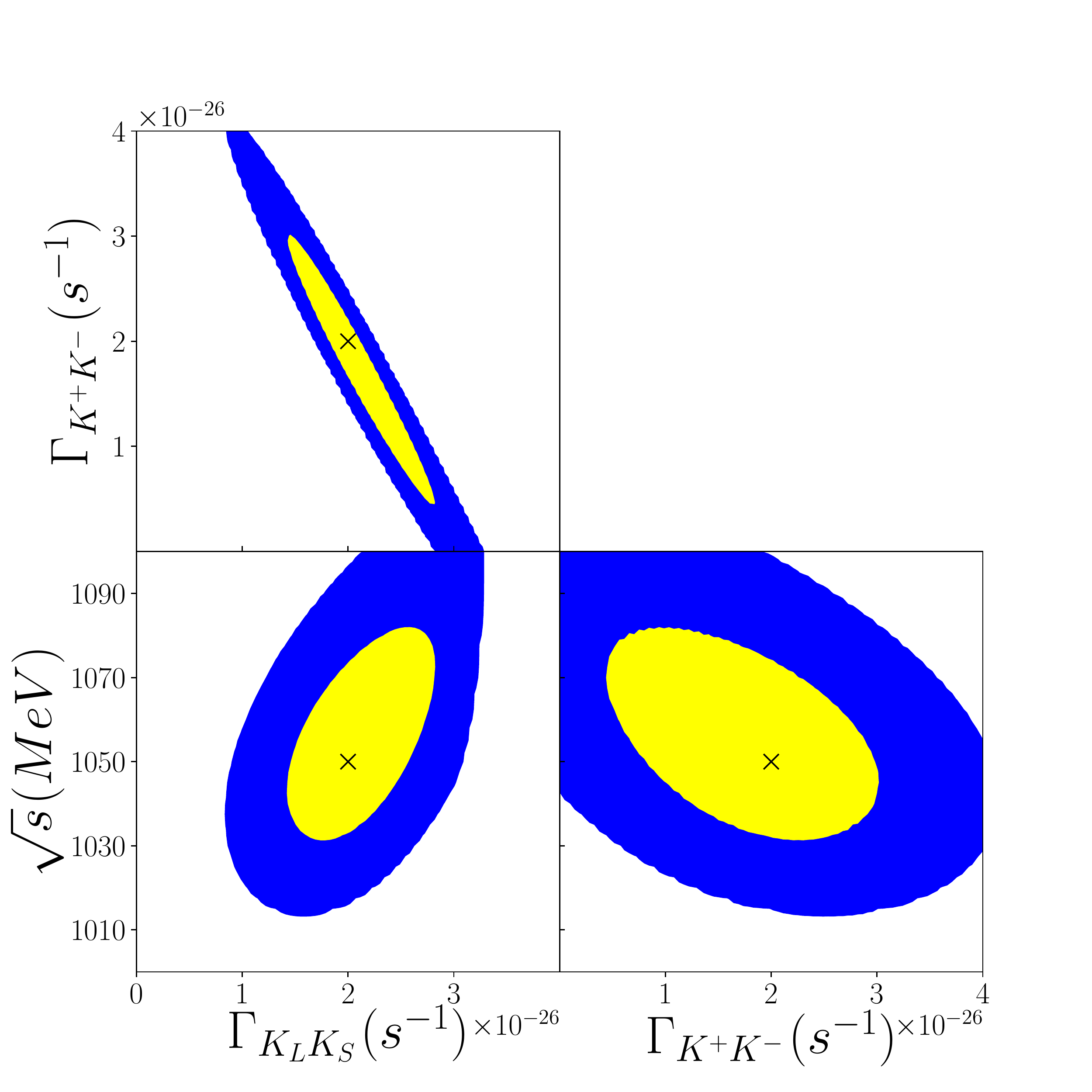}
\end{subfigure}
\begin{subfigure}[b]{0.32\textwidth}
\centering
\includegraphics[width=\textwidth]{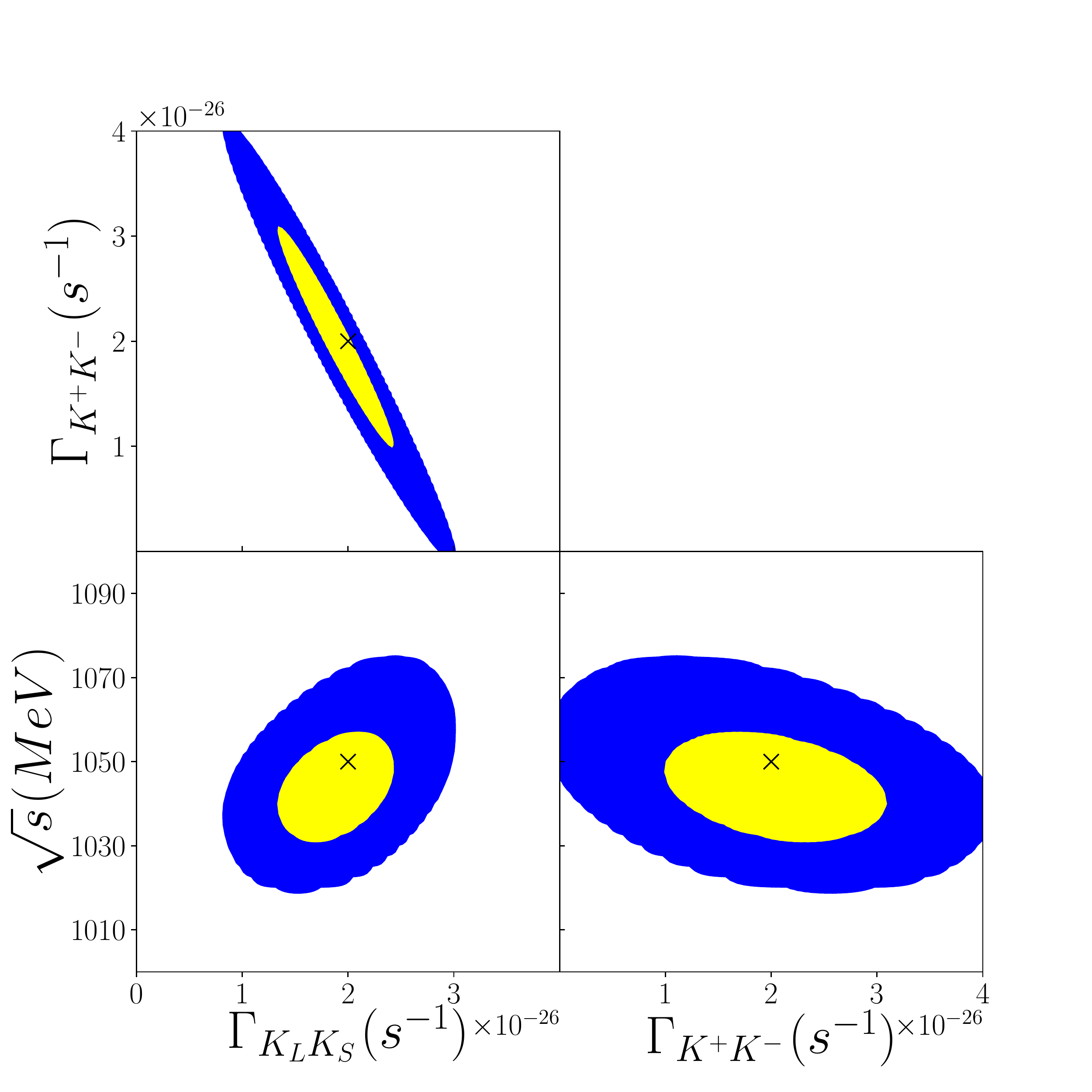}
\end{subfigure}
\caption{Projections of three-dimensional 2$\sigma$ (yellow) and $5\sigma$ (blue) confidence level contours onto 
two-dimensional subspaces, as labeled by the axes, for the $K_LK_S$ and $K^+K^-$ decay channels. The black X represents the model with which the mock data was generated. The left panel shows 3000 $\cm^2~\yr$ exposure and 30\% energy resolution. The middle panel shows 30000 $\cm^2~\yr$ exposure and 30\% energy resolution. The right panel shows 30000 $\cm^2~\yr$ exposure and 3\% energy resolution.
\label{fig:KLKSK+K__30}
}
\end{figure}

\section{Analysis and Results}
\label{sec:AnalysisResults}

We consider a mock analysis of data from 
one dwarf spheroidal galaxy: Draco.  We assume that 
it is observed by an experiment with a nominal exposure of 
$3000~\cm^2~\yr$ and an energy resolution of $30\%$.  
These approximately match the energy resolution and exposure 
expected of e-ASTROGAM, for example, with a couple of years of run time~\cite{e-ASTROGAM:2017pxr}.
We assume Draco is observed with 
an aperture of $1.3^\circ$ and that the angular resolution is small compared to this size.

For this mock analysis, we assume for simplicity that the signal 
consists of photons produced by dark matter decay,
in order to avoid the tight constraints 
on low-mass dark matter annihilation which have been obtained 
from Planck~\cite{Planck:2018vyg,Slatyer:2015jla}.\footnote{Note that the 
bounds from Planck are only stringent if dark matter 
annihilates from an $s$-wave state.  If dark matter couples to a scalar quark current, as would  
be the case if the final state were $\pi^0 \eta$~\cite{Kumar:2018heq}, 
then the dark matter annihilation cross section 
would be $p$-wave suppressed (see, for example~\cite{Kumar:2013iva}), and the Planck bounds 
would not be constraining.}  
The differential signal flux can then be expressed as 
\bea
\frac{d^2 \Phi_S}{d\Omega dE_\gamma } 
&=& \frac{\Gamma}{4\pi m_X}  J 
\frac{dN_\gamma}{dE_\gamma} ,
\eea
where $\Gamma$ is the decay rate, $dN_\gamma / dE_\gamma$ 
is the photon spectrum, and $J$ is the $J$-factor.  
We will use an angle-averaged $J$-factor for decaying dark matter 
given by~\cite{Geringer-Sameth:2014yza} 
\bea
\bar J_{dec.} &=& 5.77 \times 10^{21} \gev~\cm^{-2}~\sr^{-1} .
\eea

A detailed estimate for the astrophysical foreground and background 
in this energy range from the direction of Draco is not 
yet available.  
When next generation observatories take data, they will be 
able to estimate the backgrounds by observing slightly off axis 
from the dSph~\cite{Geringer-Sameth:2011wse,Mazziotta:2012ux,Geringer-Sameth:2014qqa,HAWC:2017mfa,Boddy:2018qur}.  But for the purpose of our mock data analysis, 
the only thing we require is an estimate of the background 
flux, and we can obtain this from data from COMPTEL and EGRET~\cite{Strong:2004de}.  
Their isotropic flux  data in the 
$0.8-30~\mev$ range can be 
fit to a power-law form, yielding a differential flux~\cite{Boddy:2015efa}
\bea
\frac{d^2 \Phi_B}{d\Omega dE_\gamma} &=& 
2.74 \times 10^{-3} \left(\frac{E_\gamma}{\mev} \right)^{-2.0} 
\cm^{-2} s^{-1} \sr^{-1} \mev^{-1} .
\label{eq:BgdFlux}
\eea
Note that a future experiment may discover many more point 
sources in this energy range, and masking these point sources 
may yield a considerably smaller flux.  In that sense, this 
estimate may be conservative.

We generate mock data by assuming a particular exposure, 
and drawing the number of signal and background photons from a Poisson 
distribution whose mean is given by the expected number 
of photons with true energy in the range $10 \mev-1 \gev$.  
The photons are assigned true energies given by the 
photon spectrum for either the background (given by Eq.~\ref{eq:BgdFlux}) or the signal model (as described in 
Sec.~\ref{sec:Spectrum}).  Finally, the measured energies 
of each photon are drawn from a Gaussian distribution 
centered at the true energy, with a width determined by 
the energy resolution.  
For this analysis, we will assume that dark matter decay 
can only yield two possible final states.  The true model 
is then defined by three parameters: the dark matter mass, 
and the partial decay width to each of the two final states.

For this analysis, we will adopt, as conservative choices, an exposure of 
$3000~\cm^2~\yr$, and an energy resolution of 
$30\%$.  But we will also consider optimistic scenarios in which the 
exposure is a factor of 10 larger ($30000~\cm^2~\yr$).
It is also interesting to consider possible improvements in energy 
resolution for upcoming experiments.  The Advanced Particle-astrophysics 
Telescope is a proposed experiment which may achieve an energy 
resolution of $\sim 10\%$ at $\sim 100~\mev$~\cite{Aramaki:2022zpw}.  Optimistically, 
we will consider an improvement in the energy resolution to $3\%$, which is a factor 
of 10 better than the conservative energy resolution.  

We then scan over models, computing the likelihood of the mock
data given the model.  In computing the likelihood, the 
combined energy spectrum of signal and background is convolved 
against the energy resolution function.  We can then 
identify the parameter point of maximum likelihood, along with 
$2\sigma$ and $5\sigma$ parameter confidence level surfaces. 
Figures~\ref{fig:PiEtaKLKS}-\ref{fig:KLKSK+K__30} show these confidence level surfaces in our three-dimensional parameter space projected onto three two-dimensional subspaces.

We first consider a scenario in which the true model is 
dark matter with a mass $m_X = 1050~\mev$, decaying to 
$\pi^0 \eta$ ($\Gamma_{\pi^0 \eta} = 2 
\times 10^{-26}~\s^{-1} $) and $K_L K_S$  ($\Gamma_{K_L K_S} 
= 2 \times 10^{-26}~\s^{-1}$).  This scenario is easily 
allowed by constraints from Planck~\cite{Planck:2018vyg}, which constrains the 
injection of energy near the time of recombination, but 
which yields bounds which are 2 orders of magnitude weaker.
In Fig.~\ref{fig:PiEtaKLKS}, we plot 2D projections of 
the $2\sigma$ (blue) and $5\sigma$ (yellow) constraint ellipsoids (the true model is denoted by a 
black X). 
In the left panel, we adopt conservative exposure and energy resolutions of 3000 $\cm^2~\yr$ and $30\%$, respectively. In the middle panel, we assume an exposure 10 times larger, with a conservative energy resolution. In the right panel, we assume an exposure 10 times larger with an energy resolution 10 times better ($3\%$).
We see that even with our conservative exposure and energy resolution,  one 
can find strong evidence for dark matter decay.   
Moreover, with the conservative exposure one can 
also determine that both the $\pi^0 \eta$ and $K_L K_S$ 
are present.  That is, the hypothesis that either channel has 
vanishing partial decay width can be rejected at 
$5\sigma$ C.L. If we assume that the branching fractions 
to the $\pi^0 \eta$ and $K_L K_S$ final states are each $50\%$ (with 
$\sqrt{s}= 1050\mev$), 
then $5\sigma$ discovery of dark matter decay can be made 
for a total decay rate as low as $\Gamma = 5\times 10^{-27}~\s^{-1}$, assuming the conservative 
exposure and energy resolution (comparable to the results of~\cite{Kumar:2018heq}).

Note, however, that because $\pi^0$ is much lighter than kaons, the available phase space 
for the $K_L K_S$ final state is only about $\sim 20\%$ of that available to the $\pi^0 \eta$ final state.  One might 
naturally expect the branching fraction to the $K_L K_S$ final state to be suppressed, making it more difficult to 
determine if this final state is present at all.  To assess this issue, we repeat the analysis above, but for 
the case in which the true model has $\Gamma_{\pi^0 \eta} = 2 
\times 10^{-26}~\s^{-1} $ $\Gamma_{K_L K_S} = 4 \times 10^{-27}~\s^{-1}$.  These results are plotted in Fig.~\ref{fig:PiEtaKLKS_phase_space_corrected}.
With a conservative choice of exposure and energy resolution, although one can easily detect the presence of the $\pi^0 \eta$ channel, 
one cannot detect the presence of the $K_L K_S$ channel.  This is not surprising, as in the absence of the $\pi^0 \eta$ channel, the 
decay rate to $K_L K_S$ alone would be so small one would not have a discovery-level detection of dark matter decay at all.  But with increased 
exposure, we see that one can detect the presence of the $K_L K_S$ at close to the $5\sigma$-level.

We next consider a true model in which the dark matter 
($m_X = 1000~\mev$) decays to the $\pi^0 \eta$ and 
$\pi \pi \eta$ states, each with partial decay width 
of $\Gamma = 2\times 10^{-26}~\s^{-1}$.  In 
Fig.~\ref{fig:PiEtaPiPiEta_30}, we again plot 
parameter constraints on this scenario, 
assuming conservative exposure and energy resolution (left panel), 
increased exposure (middle panel), or increased exposure and improved energy resolution (right panel).  
In this case, we again see that the conservative exposure and 
energy resolution is not only easily sufficient to discover the presence of 
dark matter decay, but also determine the presence of both the 
$\pi^0 \eta$ and $\pi \pi \eta$ channels.  But as with the previous case, 
we find that more optimistic choices for the exposure and energy resolution 
greatly improve parameter constraints.

Finally, we consider a true model in which dark matter 
($m_X = 1050~\mev$) decays to $K_L K_S$ and $K^+ K^-$, 
each with partial decay width $\Gamma = 2\times 10^{-26}~\s^{-1}$.
Neither of these final states produce $\eta$s through subsequent decays.
In 
Fig.~\ref{fig:KLKSK+K__30} we again plot 
parameter constraints on this scenario, 
assuming conservative exposure and energy resolution (left panel), increased exposure (middle panel), or increased exposure and improved energy resolution (right panel).
In this case, 
although the conservative choices of exposure and energy resolution are sufficient 
to detect a dark matter annihilation signal at $5\sigma$ CL, 
the presence of the 
$K^+ K^-$ channel cannot be detected at the $5\sigma$ 
level even with an exposure 10 times larger.  
However, if the energy resolution is additionally improved to $3\%$, 
then the presence of both channels can be determined with nearly $5\sigma$ confidence.

The overarching result is that, with a $30\%$ energy resolution and an exposure of 
$3000~\cm^2~\yr$, one can easily distinguish models 
which contain $\eta$s in the 
final state (with decay rates which are currently allowed and 
an ${\cal O}(1)$ branching fraction) from models which do not and 
even discriminate between different final states which contain
$\eta$s.  
But an additional improvement by a factor of 10 in both the exposure and the energy 
resolution would be needed to allow one to distinguish between different final states, 
neither of which produced $\eta$s in subsequent decays.

\subsection{Generalizations}

We have thus far considered a somewhat simple analysis, with only four final states, as a proof of 
principle.  We now 
consider if these result are expected to remain robust if we consider 
a more detailed analysis.  

For example, we have limited ourselves to final states with at most 
three mesons.  Although, for $\sqrt{s} \lesssim \gev$, one cannot have more 
than three non-pion mesons, one can potentially have several pions.  One 
expects these multi-pion states to be phase space suppressed and, thus, subdominant.
But in any case, the addition of extra pions will not affect the conclusion that 
one can readily distinguish final states containing $\eta$s from those without.  
Indeed, the main affect of adding extra pions is to reduce the kinetic energy of 
all mesons, sharpening the features in the photon spectrum around $m_\pi /2$ 
and $m_\eta /2$.

Since we have focused on the low-energy regime, we have also been able to 
assume that the only appreciable source of photons is from the diphoton decays 
of $\pi^0$ and $\eta$.  Photons also arise from the decay $\omega \rightarrow  
\pi^0 \gamma$, but this decay only has an $8\%$ branching fraction.  However, at higher 
energies, other processes can produce photons.  For example, if the $\eta'$ can be 
produced, then the decay $\eta' \rightarrow \rho^0 \gamma$ will occur with a 
$30\%$ branching fraction ($\eta' \rightarrow \gamma \gamma$ will also occur, but 
with only a $2.3\%$ branching fraction).  The decay $\eta' \rightarrow \rho^0 \gamma$ 
will produce a feature in the photon spectrum at $\sim 165~\mev$, between the features 
created by diphoton $\pi^0$ decay and $\eta$ decay.  However, as long as the center-of-mass 
energy is not too large (so the $\eta'$ is not heavily boosted), this feature should 
be narrow and readily distinguishable from the feature at $m_\eta / 2$ generated by 
diphoton $\eta$ decay.

\section{Discussion}
\label{sec:Discussion}

Thus far, we have focused on our ability to distinguish 
the mesonic final states arising from low-mass dark matter 
decay using upcoming MeV-range gamma-ray data.  We now 
address how one can use this information to learn about the 
microphysics of dark matter coupling quarks.  

For this purpose we utilize the low-energy (approximate) symmetries 
of QCD.  In particular, the mesonic final state will have the same 
$J$, $P$, $C$ and isospin quantum numbers as the quark current to which 
dark matter couples.  
Processes violating these selection rules will be suppressed by factors of 
$\alpha_{em}$, $s G_F$, or $(m_u - m_d)^2 /s$, and are expected to be small.
Thus, a determination of the final states which are 
produced by dark matter decay or annihilation can reveal the nature of 
the dark matter-quark coupling.

For example, the $\pi^0 \eta$ state is a component of an isospin triplet, is necessarily even under 
charge conjugation, and transforms under parity as $(-1)^L$, where $L$ is the 
orbital angular momentum.  
Thus, the quantum numbers of this state must be $J^{PC} = 0^{++}$ or 
$1^{-+}$.  
If it is determined that dark matter decays to 
$\pi^0 \eta$ with a nonzero partial decay width, then we would know that 
dark matter couples to a component of 
an isospin-triplet quark current with the allowed quantum numbers.  
Only the quark scalar current ($J^{PC} = 0^{++}$) satisfies this 
constraint, so the observation of a $\pi^0 \eta$ final state 
would imply that dark matter must couple to an $I=1$ scalar quark current, 
where $I$ is the isospin.

By a similar analysis, we can consider the $\pi \pi \eta$ state.  
The $\pi \pi$ state transforms as $(-1)^{L_\pi}$ under both $C$ and 
$P$, where $L_\pi$ is the orbital angular momentum of the $\pi \pi$ system.
Moreover, symmetry of the $\pi \pi$ wave function requires 
$I = L_\pi~{\rm mod}~2$.
We thus see that if the $\pi \pi \eta$ state has 
$J=0$, then it must be an isospin singlet with quantum 
numbers $J^{PC} = 0^{-+}$.  This implies that the dark matter 
couples to an $I=0$ pseudoscalar quark current.    

We thus see that, if gamma-ray telescopes provide evidence 
that dark matter decay produces both  $\pi^0 \eta$ and 
$\eta(\pi \pi)_{I=0}$ final states, one could conclude 
that the dark matter particle was spin-0, had quark couplings 
which violated $CP$, and coupled to a quark current which 
is not an eigenstate of isospin.

But if dark matter couples to scalar quark currents, and has 
sufficiently large mass, then one generically expects 
the final states $K^+ K^-$ and $K^0 \bar K^0$ to be 
produced.  Expressed in terms of mass eigenstates, the 
$K^0 \bar K^0$ state is a linear combination of 
$K_L K_L$ and $K_S K_S$, and has a photon spectrum which 
is the same as $K_L K_S$.  Thus, the presence of final states 
both with $\eta$s and also with only kaons would be evidence 
that dark matter was spin 0.

On the other hand, if dark matter is spin 1, then its decays 
cannot produce the $\pi^0 \eta$ state, although states such 
as $K^+ K^-$ and $K_L K_S$ are allowed.  If one found evidence 
of final states from dark matter decay 
containing kaons, but not of $\pi^0 \eta$, 
that would suggest that the dark matter was spin 1 or at least did 
not couple to an $I=1$ scalar quark current.

\section{Conclusion}
\label{sec:Conclusion}

We have considered the prospects for upcoming MeV-range gamma-ray 
observatories to distinguish between the hadronic final states which 
may be produced by the annihilation or decay of dark matter with 
$\sqrt{s} \lesssim {\cal O}(\gev)$.  This study is motivated by the 
fact that, for low-mass dark matter which couples to quarks, the possible hadronic final states 
are limited by kinematics and by the approximate symmetries of QCD, 
including angular momentum, $C$, $P$, and isospin.  The determination of which 
final states arise from dark matter decay or annihilation can thus provide 
information about dark matter microphysics.

For the low-mass dark matter which couples to quarks, the dominant mechanism for 
creating gamma rays is the production of $\pi^0$ or $\eta$, which decay to 
$\gamma \gamma$.  Each of these decays produces a feature 
in the photon spectrum centered at half of the meson mass.  
The spectral feature at $m_\eta / 2$ provides the most 
statistical leverage, since it is typically narrower, 
and competes against a smaller astrophysical background.

As a result, even with a $30\%$ energy resolution (as is 
expected from proposed experiments such as AMEGO and 
e-ASTROGAM), an upcoming experiment observing the Draco 
dSph with an exposure of $3000~\cm^2~\yr$ would likely be able to 
detect the presence of $\eta$s, and to distinguish 
between two possible final states containing $\eta$s.  

But if there are no $\eta$s in the final state, then it 
would be difficult, even with larger exposure, to distinguish 
if the final state is, for example, $K_L K_S$, as opposed 
to $K^+ K^-$.  But an improvement of the energy 
resolution by a factor of 10, in addition to the increased exposure, 
would allow one to distinguish between these final states.

{\bf Acknowledgements}

We are grateful to Jason Baretz for collaboration at an early stage of this 
work, and to Eric J.~Baxter for useful discussions.  
The work of J.K. is supported in part by Department of Energy Grant No. DE-SC0010504. A.R. is supported in part by the United States National Science
Foundation under Grant No. PHY-1915005.

\end{document}